# Polarization-Encoded Lenticular Nano-Printing with Single-Layer Metasurfaces

Lin Deng[1], Ziqiang Cai[1], Yongmin Liu[1,2,#]

[1]Department of Electrical and Computer Engineering, Northeastern University,

Boston, Massachusetts 02115, USA

[2]Department of Mechanical and Industrial Engineering, Northeastern University,

Boston, Massachusetts 02115, USA

[#] Corresponding author. Email: y.liu@northeatern.edu

**Abstract**

Metasurface-based nano-printing has enabled ultrahigh-resolution grayscale or color image display. However, the maximum number of independent nano-printing images allowed by one single-layer metasurface is still limited despite many multiplexing methods that have been proposed to increase the design degree of freedom. In this work, we substantially push the multiplexing limit of nano-printing by transforming images at different observation angles into mapping the corresponding images to different positions in the Fourier space, and simultaneously controlling the complex electric field across multiple polarization channels. Our proposed Polarization-Encoded Lenticular Nano-Printing (Pollen), aided by a modified evolutionary algorithm, allows the display of several images based on the viewing angle, similar to traditional lenticular printing but without requiring a lenticular layer. In addition, it extends the display capability to encompass multiple polarization states. Empowered by the ability to control the complex amplitude of three polarization channels, we numerically and experimentally demonstrate the generation of 13 distinguished gray-scale Chinese ink wash painting images, 49 binary patterns, and three sets of 3D nano-printing images, totaling 25 unique visuals. These results present the largest number of recorded images with ultra-high resolution to date. Our innovative Pollen technique is expected to benefit the development of modern optical applications, including but not limited to optical encryption, optical data storage, lightweight display, and augmented reality and virtual reality.

**Introduction**

Lenticular printing, which utilizes lenticular lenses to create printed images that can appear to change or move at different viewing angles, has been widely used in arts, marketing, and advertising in the past decades[1-4]. For conventional lenticular printing, multiple sets of segmented images are combined behind a lenticular grating consisting of a microlens array as shown in **Figures 1a** and **1b**. As the viewing angles are altered, the corresponding images projected to the viewer's eyes are sequentially switched. This allows for the switching between different images, as well as 3D prints that exhibit the three-dimensional effect of the target object by inserting 2D images from a different perspective. Nevertheless, lenticular printing has some drawbacks and cannot be directly applied to nano-optics. It requires top-layer grating microlenses to deflect light, and all target images need to be divided into many small pieces and then assembled at the bottom of the grating, which inevitably causes the discontinuity of the images. These prerequisites also make it difficult for this technology to be directly scaled down to the nanoscale and used in metasurface nano-printing.

On the other hand, metasurfaces, which are artificial two-dimensional nanostructures, have emerged as a new frontier in optics and photonics because of their superior ability to control the amplitude, phase, and polarization of the scattering light with high precision. Many innovative functions using metasurfaces, such as metalens[5-8], holographic image display[9-15], vortex beam generator[16-19], and beam steering and shaping[20, 21], have been reported. Nano-printing, which refers to the technique that can display grayscale or color images at the surface of the device, is also one of the most important applications of metasurfaces[22-32]. For the conventional nano-printing technique, the image observed on the surface of the device is directly related to the amplitude response of the scattering light after the incident wave interacts with the nanostructures. There are mainly two methods to realize nano-printing[22]. One is to manipulate the polarization-assisted intensity based on Malus law, and the other one is to control the spectrum of the scattered light based on structured color. For the simplest case, we expect metasurface devices to exhibit one nano-printing image with the given monochromatic light input at a certain input-output polarization pair according to Malus's law[23, 29]. When non-orthogonal polarization states are considered in the design, two independent nano-printing images can be integrated into one metasurface based on Malus law.[28, 33, 34] Recently, researchers have also demonstrated that by manipulating the deflection angle of coherent pixels[35], up to 3 nano-printing images can be observed by utilizing various combinations of light properties such as wavelength, polarization, and angle of incidence. Some recently published work also shows that with the precise control of the Jones matrix, 2 and 3 independent nano-printing images at multiple polarization channels can be realized[36-38]. Despite the extensive effort to increase the information capacity of nano-printing, the total amount of images that can be displayed by a single-layer metasurface remains significantly limited, so far up to 3. Therefore, a new design strategy is urgently required to further expand the multiplexing capacity of nano-printing.

Therefore, it is of great interest to develop a novel display technique that inherits the unique properties of metasurface-based nanoprinting and lenticular printing while overcoming their limitations. To be more specific, as shown in **Figure 1c**, a single-layer metasurface device that can display various independent nanoscale images with subwavelength pixel size when viewed from multiple different angles, with no requirement of micro-lens as well as the segmented image region. In three separate polarization channels, patterns corresponding to specific categories can be observed at different viewing angles. In this case, these include mathematic symbols, Latin characters, and Chinese strokes. We would expect that the introduction of a novel design method capable of further reducing image pixels to the nanoscale, combined with the ability to fully manipulate multiple polarization channels, will significantly increase the information capacity and elevate the total number of permissible nano-printing images to unprecedented levels.

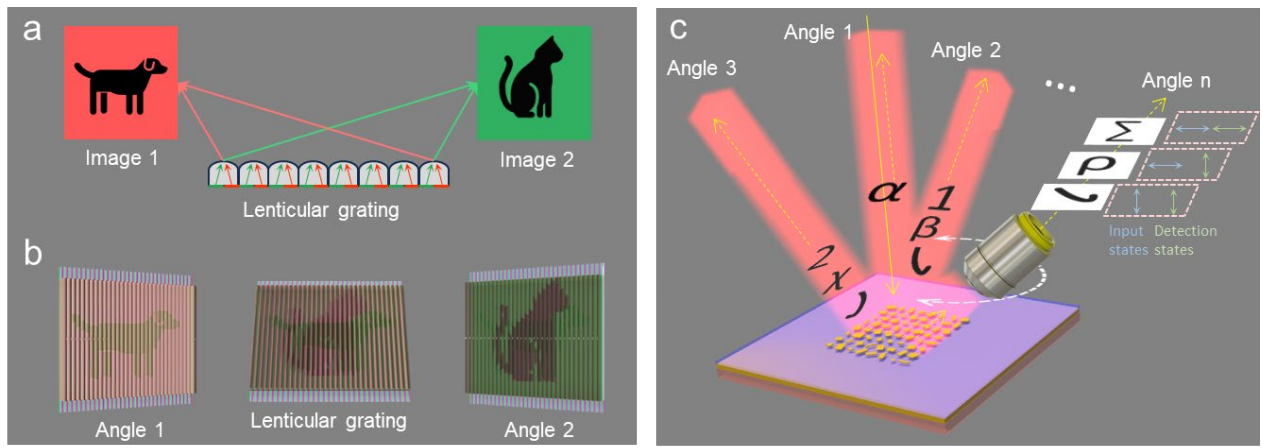

Figure 1. Comparison of different methods that can show nano-printing images. **a**, Illustration of the conventional lenticular printing method where two sets of images are interleaved under cylindrical lenses. Then two images, which are represented by the two pure-color images of a cat and a dog for simplicity, can be observed at two viewing angles. **b**, 3D illustration depicting the display of two images based on the respective viewing angles using a lenticular grating. **c**, Illustration of our method that allows the independent display of multiple nano-printing images corresponding to the various viewing angles as well as the polarization channels with the help of our single-layer lenticular metasurface.

In this work, we demonstrate a novel Pollen technique, that can show up to 49 binary images and 13 grayscale Chinese ink wash painting images at different viewing angles and polarization states. our method only requires a single-layer metasurface without the need for cylindrical lenses, and all generated images are free from discontinuous stripes or crosstalk. The working wavelength is set at 1064 nm and the pixel of each image is only $600\ nm \times 600\ nm$, corresponding to a resolution of 42.3 k dots per inch (dpi). To the best of our knowledge, this marks the first demonstration of the highest multiplexing capacity on a single metasurface device for both grayscale and binary nano-printed images while maintaining ultra-high dpi. A detailed comparison of our work with published works can be found in **Section 1** of the **supplementary materials**.

**Results and discussion:**

The display of the nano-printing images is directly related to the amplitude information of the diffracted light reflected by the metasurface plane. To better explain the working principle of our design, we begin by describing the complex electric field of periodic structures on a certain plane under a specific polarization state and wavelength, which is given by

$$E(x,y) = A(x,y)\exp(j\phi(x,y)) \tag{1}$$

Here $A(x,y)$ and $\phi(x,y)$ represent discrete arbitrary amplitude and phase distribution of the pixel at position $(x,y)$. The periodicity of the structure along the $x$ and $y$ directions are $P_x$ and $P_y$. We can always rewrite this field as a summation of multiple complex electric fields $E_i$,

$$E_{tot}(x,y) = \sum_{i=1}^{n} E_i(x,y) = \sum_{i=1}^{n} A_i(x,y)\exp(j\phi_i(x,y)) \tag{2}$$

If we assign $A_i$ with predefined amplitude distribution of the nano-printing images $NP_i$, and add a planar wave wavefront with the wavevector of $(k_{ix}, k_{iy})$ to the phase term, then the required $E(x,y)$ can be rewritten as

$$E(x,y) = E_{tot}(x,y) = \sum_{i=1}^{n} NP_i(x,y)\exp(j(k_{ix}x + k_{iy}y)) \tag{3}$$

Equation (3) indicates that the beam propagates along the corresponding direction while carrying the information of the nano-printing images at the metasurface plane. Then the electric field in the k-space can be rewritten by discrete 2D Fourier transform as

$$E(k_x, k_y) = \mathcal{F}(E(x,y)) = \sum\sum E(x,y)\exp\left[-j\left(\frac{2\pi k_x}{M}x + \frac{2\pi k_y}{N}y\right)\right] \tag{4}$$

where M and N are the dimensions of the 2D grid.

Replacing $E(x,y)$ with Equation (3), we can obtain,

$$E(k_x, k_y) = \sum_{i=1}^{n}\left(\sum\sum NP_i(x,y)\exp\left(-j\left(\left(\frac{2\pi k_x}{M} - k_{ix}\right)x + \left(\frac{2\pi k_y}{N} - k_{iy}\right)y\right)\right)\right) \tag{5}$$

Equation (5) shows that after performing the Fourier transform to $E(x,y)$, for each decomposed $E_i(x,y)$, its corresponding electric field in the k-space is the Fourier transform of the amplitude distribution centered at $(k_{ix}, k_{iy})$. This provides us an opportunity to separate all the wave component $E_i$ in the k-space without crosstalk if appropriate amplitude profile $NP_i$ and wavevector $(k_{ix}, k_{iy})$ are predefined in the design. Although the separation of the k-space is in the far field, it can affect the images collected by objective lenses at the surface of the device. In the following, we will demonstrate how the splitting of the wavevectors can help with the design of displaying multiple images via one single-layer metasurface device.

As an example shown in **Figure 2a**, we define an amplitude profile consisting of four waves with distinct grayscale Chinese ink wash paintings of fishes while setting specific propagating angles in four different directions. Based on Equation (5), we have,

$$E(k_x, k_y) = \sum_{i=1}^{4}\left(\sum\sum NP_i(x,y)\exp\left(-j\left(\left(\frac{2\pi k_x}{M} - k_{ix}\right)x + \left(\frac{2\pi k_y}{N} - k_{iy}\right)y\right)\right)\right) \tag{6}$$

Then, the Fourier transform of those beams will be located around the four predefined wavevectors as presented in **Figure 2b**. As we know, the numerical aperture of a lens denotes the maximum angle of the diffracted light that the lens can collect. To observe the nano-printing images, a lens with a suitable numerical aperture, which only collects the information in the desired region of the k-space while filtering out other components, is applied to directly view the device at the corresponding observation angle. To make sure that we can only see the target images without any crosstalk from other channels when observing at the corresponding angles, and to enable clear observation of the displayed image with appropriate magnification, we chose a 10× objective lens with an NA of 0.25. As shown in the left panel of **Figure 2b**, the red circles represent the angular spectrum region allowed to be collected by the lens and it can cover most of the desired Fourier component. After applying inverse fast Fourier transform (IFFT) only for the k-components inside the red circles, we can retrieve the nano-printed images as shown in the right panel of **Figure 2b**, which agree well with our target patterns. It is important to emphasize that, in contrast to holographic images, the angle-multiplexed nano-printed images exist on the same surface and can solely be differentiated with the correct observation angles.

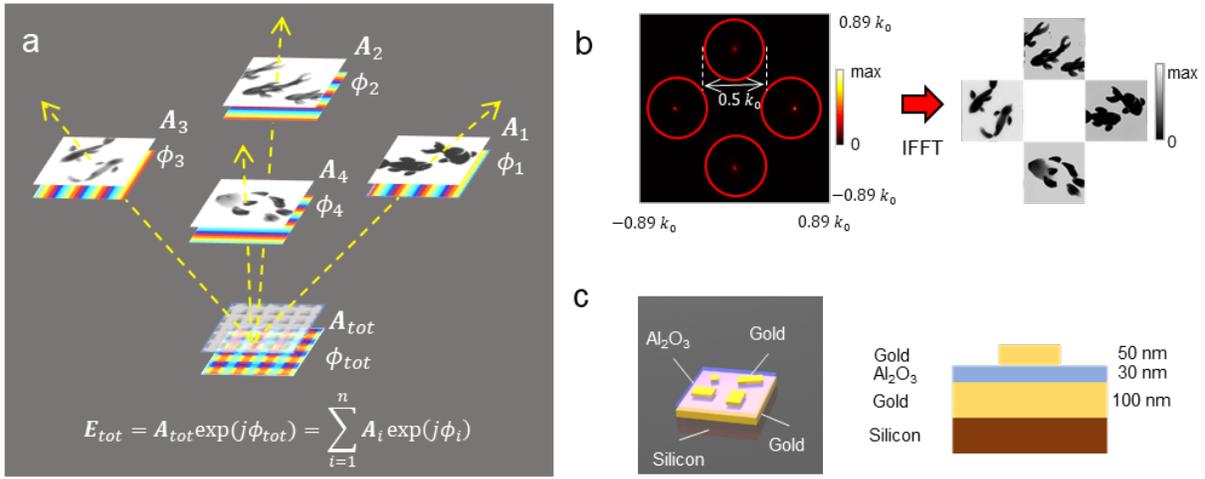

Figure 2. **a**, Illustration of combining the four waves with different amplitude distributions and wavevectors into one electric field. The electric field of the metasurface region is zoomed in for a better visual effect. **b**, Left panel: Fourier transform of the combined electric field. The red circles denote the collecting range of the objective lens with the numerical aperture of 0.25 when placed at different viewing angles. Right panel: recreated nano-printing images generated from the IFFT of the k-components inside the four red circles. **c**, Schematic of the unit cell or the pixel of the metasurface. Four gold nanodisks with different lengths, widths, and rotation angles are on top of a 30 nm $Al_2O_3$ layer, with 100 nm gold film and the silicon substrate at the bottom.

Although the above lenticular nano-printing method is based on one polarization state and one wavelength, it can be directly extended to multiple channels when more design degrees of freedom are involved. For example, we can apply our method to three polarization channels, which is the upper limit of independent channels without crosstalk for a single metasurface[37, 41], and achieve 13 grayscale Chinese ink wash paintings in total. For each of the co-polarization channels (i.e., $E_{xx}$ and $E_{yy}$), the total electric field is the combination of four beams with four nano-printing images and wavevectors. For the cross-polarization channel (i.e., $E_{xy} = E_{yx}$), since there is no crosstalk between the incident and reflected light, an additional Chinese ink wash painting can be added for the viewing angle perpendicular to the metasurface. Therefore, five beams will be summed together to create $E_{xy}$.

As shown in **Figure 2c**, our metasurface is based on a metal-insulator-metal (MIM) structure. Each pixel of our design consists of four gold nanodisks, in which the length, width, and rotation angles are optimized by a modified genetic algorithm, allowing us to well control the complex amplitude of the three Jones elements[19, 41]. The details of the genetic algorithm can be found in **Section 2** of the **Supplementary Materials**. Restrictions about the maximum size of each nanodisk as well as the gap distance between the neighboring nanodisks are applied in our designing process so that the size of each pixel of the image is set at 600 $nm$ × 600 $nm$, which is much smaller than the working wavelength 1064 nm. According to angular spectrum theory, the maximum range in the k-space that we can realize for the complex amplitude control is $(-0.83\ k_0, 0.83 k_0)$ as shown in **Figure 2b**. More discussion about the shrink of the effective k-space with coherent supercells can be found in **Section 3** of **Supplementary Materials**. We have also deliberately designed the wavevector of the four angles so that there is no crosstalk from other channels or reflected light when observing with the NA=0.25 lens. The propagation direction of the four waves is set at (-34, 0), (0, 34), (34, 0), and (0, -34) degrees, corresponding to the wavevector of $(-0.56\ k_0, 0)$, $(0, 0.56\ k_0)$, $(0.56\ k_0, 0)$, $(0, -0.56\ k_0)$. The challenge of displaying multiple nanoprinting images can now be simplified as the separation of wavevector distributions in the k-space, where the objective lens acts as a spatial bandpass filter. In this way, we can realize a single-layer Pollen device that shows images in multiple viewing angles and does not require lens grating or large-size assembled pixels. We have also performed the full-wave simulation to verify our method. The details as well as the experiment results of the 13 zodiac sign images can be found in **Section S4** of the **Supplementary Materials**.

The experimental setup is illustrated in **Figure 3a**. The monochromatic laser with a wavelength of 1064 nm is illuminated perpendicular to the structure surface after passing through one half-wave plate, one linear polarizer, and one focusing lens. Then the light scattered from the metasurface is collected by a 10× objective lens with an NA of 0.25 at different angles. Finally, light is projected to a CCD camera after passing through another polarizer and lens. To ease the measurement, we fix the objective lens and rotate the sample and the polarizers correspondingly. **Figure 3b** presents the SEM images of the fabricated metasurface, which is 219.6 $\mu m$ × 219.6 $\mu m$ in size. The zoomed-in SEM image confirms the good quality of the fabricated nanodisks. Due to equipment constraints, we were unable to fabricate the metasurface in a single process. Hence, we stitched four separate sample pieces, each with a size of 109.8 $\mu m$ × 109.8 $\mu m$, to make the final device. More discussion about the robustness comparison of the holographic images and nano-printing images regarding the misalignment can be found in **section S5** of the **supplementary materials**. Although some dislocation can be spotted at the boundaries of the subregions in the SEM image (shown in **Figure S6** in the **supplementary materials**), the tested nano-printing images still have high quality despite two crossed black lines with a width of $1 - 2\ \mu m$ are shown at the stitching boundaries.

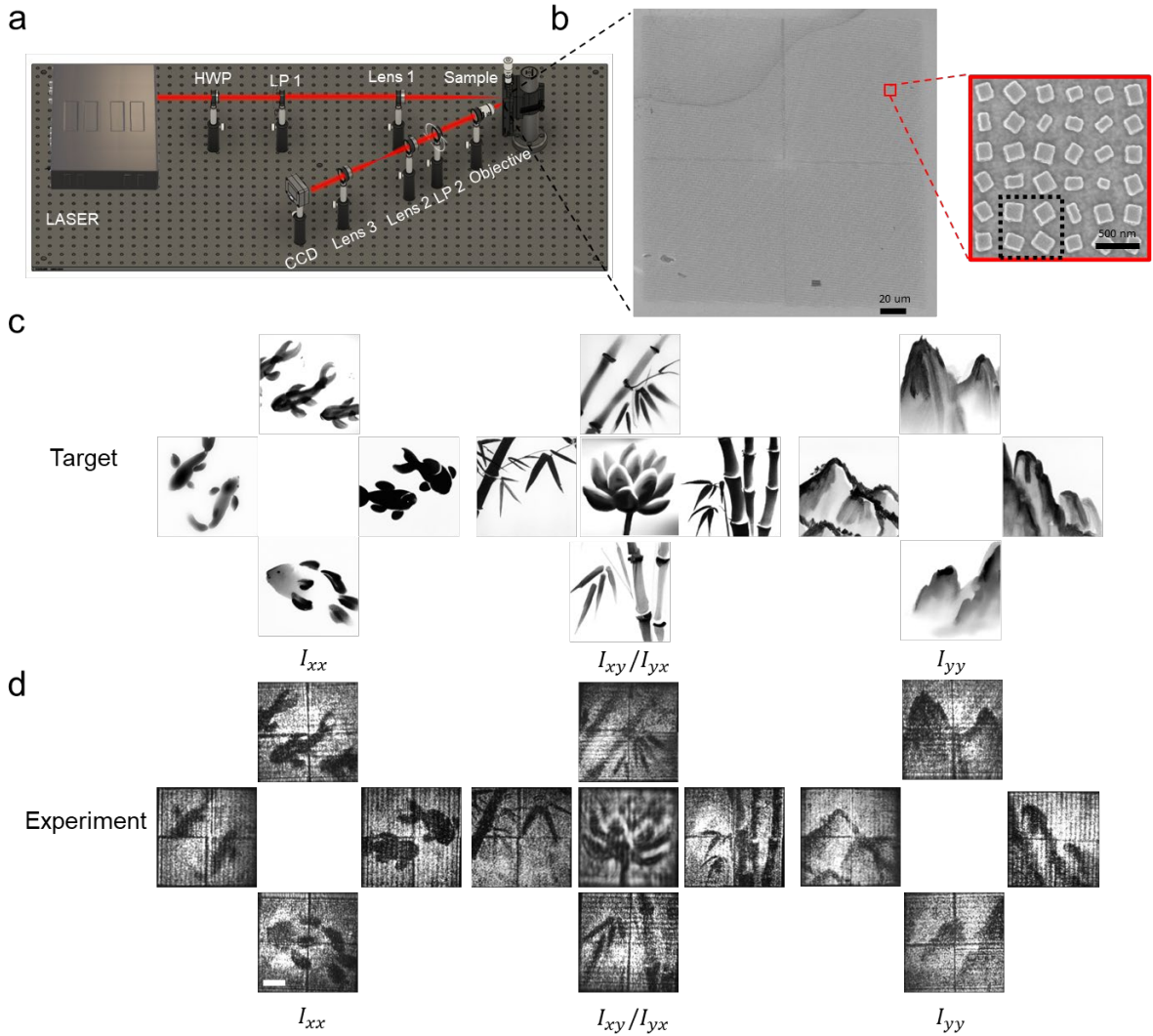

Figure 3. **a**, Optical measurement setup of recording the nano-printing images. One 10 × objective lens with NA = 0.25 is used in the experiment to observe the metasurface plane at different angles. The wavelength of the illumination is 1064 nm. HWP: half waveplate. LP: linear polarizer. **b**, SEM image of the fabricated metasurface. The black box denotes one pixel of the nano-printing images. **c**, Target and **d**, experiment results of grayscale nano-printing images for different viewing angles and polarization states. The recorded images are converted to grayscale images for better visual effects. Scale bar: 40 $\mu m$.

Figure 3c and 3d show the targets and experimental results of grayscale nano-printing images for different polarization states and observation angles. The images include four paintings of fish, four of bamboo, and four of mountains. Furthermore, at the zero-degree reflection angle, an additional lotus pattern can be observed in the cross-polarization channel. The experimental results all match well with our targets. Some details, such as the gradual grayscale transition f, the fish fins, and the leaves of the bamboo are distinctly observable. It is noted that observing the zeroth-order diffraction necessitates the use of a beam splitter, which increases the distance between the objective lens and the metasurface. To accommodate this, we switched to a 4 × objective lens with a longer focal distance, which to some extent compromised the image quality. Since the dimension of the minimal pixel in our design is 600 nm × 600 nm, the image resolution of our device can reach 42.3 k dpi. It represents an increase of 32 times compared to traditional image display methods that rely on liquid crystal technology, considering the pixel dimensions of a standard liquid crystal display is around 19 $\mu m \times$ 19 $\mu m$. This advancement stands as a substantial enhancement in terms of device miniaturization.

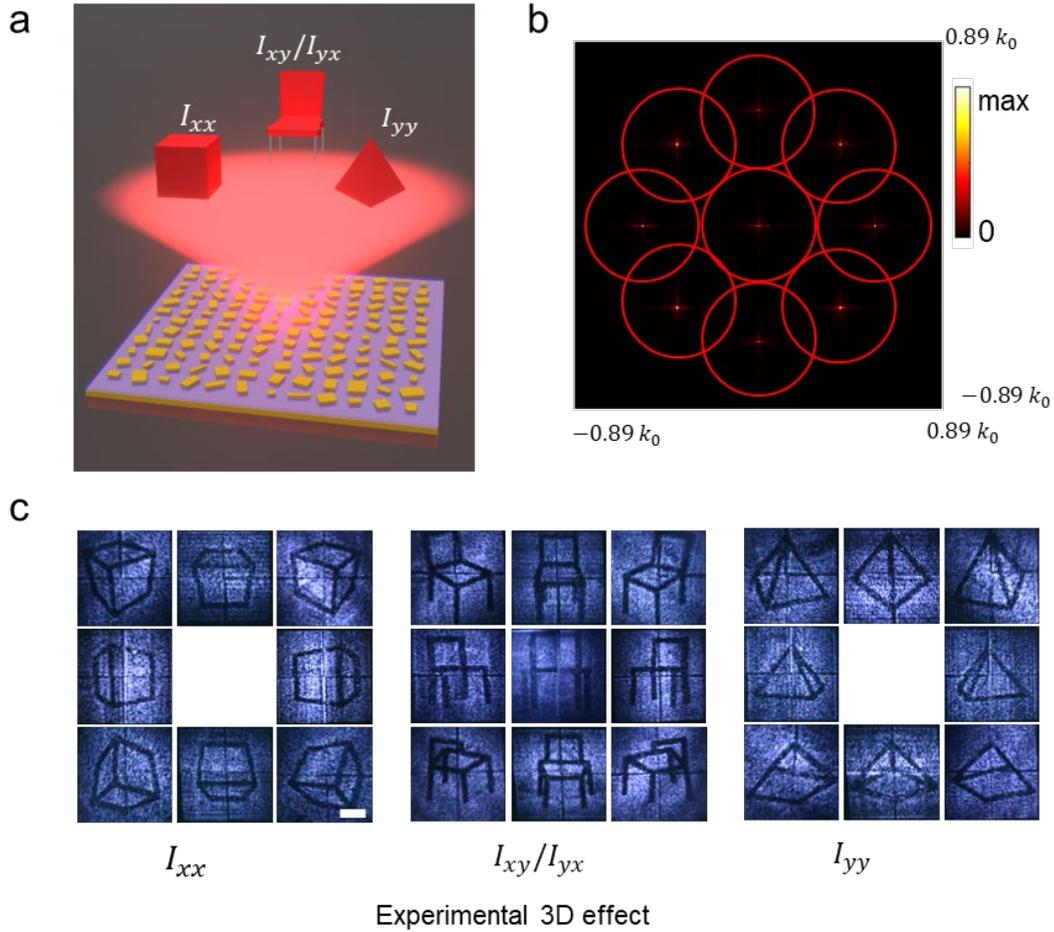

Figure 4. **a**, Illustration of the metasurface showing 3D lenticular nano-printing images. **b**, Fourier transform of the combined electric field. The red circles denote the collecting range of the objective lens with an NA of 0.25 when placed at different viewing angles. **c,** experimental results of 3D lenticular nano-printing images. For $xx, xy, yy$ polarization channels, the 3D effect of three objects including a cube, a chair, and a tetrahedron, can be observed at eight or nine viewing angles. Scale bar: 50 $\mu m$.

Akin to 3D lenticular images that can show the effect of three-dimensional depth of the object, our Pollen can also produce stereoscopic effects of the target objects by deliberately setting 2D images corresponding to the viewing angles. As shown in **Figure 4a**, we can realize the 3D effect of multiple objects, such as a cube, a chair, and a tetrahedron, from the corresponding perspective at three polarization states. To make our 3D lenticular nano-printing images more realistic, we will further expand the display channels, encompassing a total of eight wavevectors along the horizontal, vertical, and diagonal directions. For the cross-polarization channel, we can add an additional viewing angle along the surface's normal direction, similar to the first example. As shown by the wavevector diagram in **Figure 4b**, although there are overlapping regions between the nine collected regions in the k space, the majority of the Fourier components are still confined inside each region, which indicates the crosstalk between different channels can be neglected.

As we further expand the viewing angles to 8 or 9, the number of allowable nano-printing images increases to 25. From the experimental results shown in **Figure 4c**, nano-printing images display the outlines of these objects when observed from different angles. A clear transition between the cube and chair can be observed in **Supplementary Movie 1**. Such a device offers a visually striking and engaging effect with depth and motion without the need for 3D glasses, which could potentially benefit the development of modern augmented reality, virtual reality, optical encryption, advertising, and education.

Moreover, as discussed in **Section S6** of the **supplementary materials**, the major parts of the initial images are confined in the low-frequency domain of the k space, which offers us the opportunity to further increase the display channel by shrinking the NA of the objective lens and integrating more observation angles. As shown in **Figure 5a**, with an objective lens of the NA = 0.1, we can set two collecting angles, corresponding to 0.3 $k_0$ and 0.6 $k_0$, respectively, along the radial direction with a shrunk NA range. As a result, 49 nanoprinting images that show 17 Latin letters, 16 mathematic symbols, and 16 Chinese strokes can be integrated into a single metasurface as shown in **Figure 5c**. The experimental findings exhibit satisfactory agreement with our objective, despite a reduction in the signal-to-noise ratio caused by the inherent decrease in intensity within each channel as the number of channels is increased. The nanoprinting images captured through the cross-polarization channel demonstrate enhanced contrast compared to the co-polarization channels, attributed to the presence of reduced background noise levels. In practical information encryption applications, when both the sender and receiver have access to the specific custom key pairs, we can obtain corresponding Chinese characters or sentences, formulas, or even equations by combining the patterns of these three channels using a prearranged sequence as shown in **Figure 5d**. Alternatively, we can encrypt the corresponding text information into a combination of specific polarization and angular information. To the best of our knowledge, this marks the highest capacity reported in metasurfaces nanoprinting featuring polarization and angle multiplexing designs. We also showcase an increase in the number of images to 61 by sacrificing the signal-to-noise ratio of the displayed image. More details on this trade-off are provided in **Section S6** of the **supplementary materials**. Furthermore, by integrating with complementary methods such as wavelength multiplexing, this approach can be refined to boost capacity without substantially impacting image quality, thus enabling the presentation of vivid nanoprinting images.

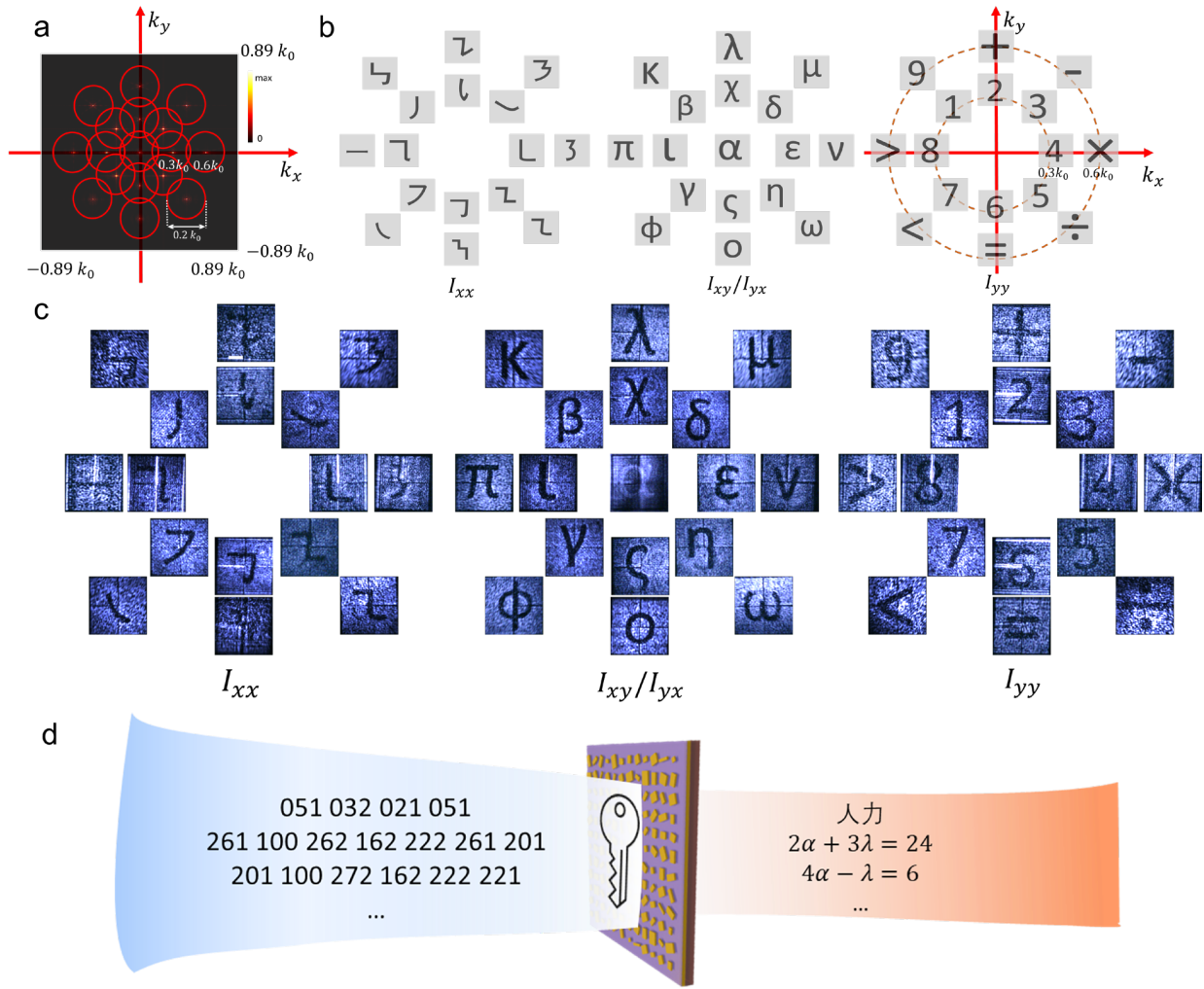

Figure 5 **a**, Fourier transform of the combined electric field that can show up to 49 nanoprinting images. The red circles denote the collecting range of the objective lens with the numerical aperture of 0.1 when placed at different viewing angles. **b**, Target patterns, and **c**, experimental results of lenticular nano-printing images. **d**, Illustration of information encryption enabled by Pollen. In the left part, each set of three numbers corresponds to specific polarization channels: 0 represents polarization channel xx, 1 represents polarization channels xy/yx, and 2 represents polarization channel yy. The second number in each set indicates the order in the clockwise azimuthal direction, commencing from the positive $k_x$ direction as 0. The third number denotes the order in the radial direction, commencing from the center as 0.

**Conclusion:**

To summarize, we demonstrate polarization-encoded lenticular nano-printing (Pollen) with single-layer metasurfaces. With the help of an inverse design algorithm, we successfully realize 13 independent gray-scale nano-printing images. We also demonstrate the display of 3D lenticular nano-printing images that can create the 3D effect for three objects including a cube, a chair, and a tetrahedron in all three polarization channels consisting of 25 images. When observed with a small NA lens, up to 49 independent nanoprinting images can be displayed, which is the maximum number of nano-printing images allowed by a single metasurface to the best of our knowledge. Owning to the subwavelength pixel size of $600\ nm \times 600\ nm$, the resolution of the nanoprinting image can reach 42.3k dpi. Our design also shows great robustness for stitching errors when the fabrication of large-size devices is required. With the help of other phase retrieval functionals such as the GS algorithm[43, 44], we can further expand the information channel by adding holographic images to the k-space in the corresponding observation directions. We also expect that with this novel multiplexing framework, we can further expand the information channel with other available design degrees of freedom such as wavelength multiplexing[45-47], which enables the display of vivid color images. We believe that our work paves the way for achieving parallel processing, low crosstalk, and high-capacity optical display, as well as optical encryption and information storage systems. These advancements have the potential for widespread applications in the future.

**Method:**

Fabrication: First, 50 nm-thick Au was deposited on the top of a Si wafer by an e-beam evaporation process. After cleaning the wafer with acetone and isopropanol, a layer of $Al_2O_3$ with 30 nm thickness was deposited using the atomic layer deposition (ALD) process. Then, the 50 nm Au metasurface was fabricated by e-beam lithography, followed by the e-beam evaporation and lift-off process.